# Rewards and the evolution of cooperation in public good games

Tatsuya Sasaki[1,2*] and Satoshi Uchida[3]

[1]Faculty of Mathematics, University of Vienna, Oskar-Morgenstern-Platz 1, 1090 Vienna, Austria

[2]Evolution and Ecology Program, International Institute for Applied Systems Analysis, Schlossplatz 1, 2631 Laxenburg, Austria

[3]Research Center, RINRI Institute, Misaki-cho 3-1-10, Chiyoda-ku, 101-8385 Tokyo, Japan

[*]Corresponding author (tatsuya.sasaki@univie.ac.at)

October 20, 2013

**Summary:** Properly coordinating cooperation is relevant for resolving public good problems such as clean energy and environmental protection. However, little is known about how individuals can coordinate themselves for a certain level of cooperation in large populations of strangers. In a typical situation, a consensus-building process hardly succeeds due to lack of face and standing. The evolution of cooperation in this type of situation is studied using threshold public good games in which cooperation prevails when it is initially sufficient, or otherwise, it perishes. While punishment is a powerful tool to shape human behaviours, institutional punishment is often too costly to start with only a few contributors, which is another coordination problem. Here we show that whatever the initial conditions, reward funds based on voluntary contribution can evolve. The voluntary reward paves the way for effectively overcoming the coordination problem and efficiently transforms freeloaders to cooperators with a perceived small risk of collective failure.

# 1. Introduction

Public goods, such as clean energy and environment protection, are building blocks of sustainable human societies, and those failures can have far-reaching effects. Private provision of public goods, however, poses issues where cooperation and coordination often do not succeed (e.g., [1]). First, voluntary cooperation to provide public goods suffers from self-interest behaviours. Exploiters can freeload on the efforts of others. In collective actions, proper coordination among individuals is usually required to attain a cooperation equilibrium. Otherwise, the advantage of freeloading leads individuals to end up being stacked with the non-cooperation equilibrium, which is a social trap.

The coordination problem has been broadly studied by game theory, and its ubiquity is indicated by a variety of names: coordination game, assurance game, stag-hunt game, volunteer's dilemma, or start-up problem [2–4]. Evolutionary-game models tackling sizeable groups often are built on public good games of cooperation and defection, but have generally resulted in a system that has a couple of equilibria (ones with no cooperation and certain-level cooperation) [5,6]. It has been a conundrum to develop a mechanism that allows populations independent of the initial conditions to evolve towards the cooperation equilibrium. Especially in a case where unanimous agreement is required for the public good, the situation is most stringent, with the only desirable initial condition being a state in which almost all cooperate. Theoretical and empirical analyses have clarified that prior communication [7] or social exchange situation [8] can facilitate selection of the cooperation equilibrium. Little is known, however, about how equilibrium selection can materialize from one-shot anonymous interactions in large populations, in which such a consensus-building process is less likely to succeed. Previous studies showed that the higher the risk perception of collective failure, the higher the chance of coordinating cooperative actions [9–11]. Recent research has put forth that considering institutional punishment can further relax the initial conditions for establishing cooperation [12].

What happens if one considers reward, instead of punishment? Reward is one of the most-studied structural solutions for cooperation in sizeable groups, and better inspires cooperation [13,14]. While in real life there exist a huge array of subsidy systems for encouraging cooperative actions, here we turn to endogenous fundraising. (See [15] for formal rewards.) Early work revealed that replicator dynamics [16], whereby the more successful strategy spreads further, can lead to dynamic maintenance of cooperation in public good games with reward funds [17]. This model considered three strategies: to be (i) a cooperator or (ii) defector in the standard public goods game, or to be

(iii) a rewarder that contributes both to the public good and to the reward fund. Only those who contribute to the public good are invited into sharing the returns from the reward fund. The rewarders can spread even in a population of defectors, since these persons are excluded from the rewards. The fundraising itself, however, is voluntary and costly. This incentive scheme thus can easily be subverted by "second-order-freeloading" cooperators who contribute to the public good, but not to the rewards. In the next step, since contribution to the public good is also costly, cooperators will be displaced by "first-order-freeloading" defectors. This leads to a rock-scissors-paper type of cyclical replacement among the three strategies.

**2. Model**

We extend public good games with reward funds [17] with a provision threshold [12], which can easily materialize a coordination situation. We consider infinitely large, well-mixed populations, from which $n$ individuals with $n > 2$ are randomly sampled and form a gaming group. After one interaction, the group is dissolved. We assume the three strategies as before: both the rewarder and the cooperator are willing to contribute at a cost $c > 0$ to itself; the defector contributes nothing and thus incurs no cost. 100% of the public benefit is provided only if the number of contributors $m$ ($0 \leq m \leq n$) in the group exceeds a threshold value $k$ ($1 \leq k \leq n$); or otherwise, part of the public benefit, discounted by a risk factor $p$ ($0 \leq p \leq 1$), is provided. However, the resulting benefit equally goes to every player, whatever she/he contributes. The individual benefit is given by $B(m) = 1 - p$, if $m < k$, or otherwise, $B(m) = 1$ (figure 1).

Next, we consider a voluntary reward fund for the threshold public good game. Beforehand, only the rewarders are willing to spend $c' > 0$ to the fund; then, after the game, the integrated fund multiplied by an interest rate $r' > 1$ will be shared equally among the $m$ contributors ($i$ rewarders and $m - i$ cooperators) to the public good. The reward fund is thus a "club" good, excluding the defectors. In sum, a rewarder earns $B(m) - c + c'r'i/m - c'$, a cooperator, $B(m) - c + c'r'i/m$, and a defector, $B(m)$.

**3. Results**

First, we look at the evolutionary outcomes without the rewarders. The step function $B(m)$ with the intermediate threshold value $k$ (figure 1) can lead to bistability of both no cooperation and a mixture of cooperation and defection for the sufficiently large risk factor $p$ for $1 < k < n$, and for $k = 1$, with only the mixed one [10,18]. Pure coordination between no and 100% cooperation occurs if and only if $k = n$, a case where avoiding the collective failure requires homogeneous cooperation among all participants. It holds for $1 < k \leq n$ that the larger the risk $p$, the smaller (larger) the basin of attraction for a no (certain) cooperation equilibrium [10]. And in particular, to bring about bistability the critical risk factor $p^*$ takes its smallest value in the case of the unanimous agreement.

The evolutionary outcomes change dramatically with the rewarders (figure 2a–e). The analytical investigation shows that if rewards are considered, the replicator dynamics leads the population to escape the non-cooperation equilibrium (D) and then evolve to the mixed equilibrium ($X_2$) for $1 < k < n$, which for $k = n$ is 100% cooperation (C), once a bistable situation for the dynamics between cooperation and defection arises. For a certain level of rewards, first, no cooperation is negated by the spread of the pro-social, rewarding cooperators. The rewarders dominate the defectors as long as the most promising return of the fund $c'(r' - 1)$ is greater than $c$, a cost for the public good. Second, the temptation to withhold contribution to the reward pool tends to downgrade the pro-social efforts: the non-rewarding cooperators then subvert the population of the rewarders (R), and will spread over the population. This is common for whatever $p$ and $k$, and leads to forming a roundabout to the side of the cooperation equilibrium C. In the absence of bistability for the threshold public good game, then the population state pulls back to the non-cooperation D's side. The population can end up with complex dynamics, such as boundary orbits traveling the three homogenous state in rotation of D→ R→ C→D, or an interior limit cycle (figure 2b). Similar oscillatory dynamics for cooperation and rewards have been obtained in models more complicated by reputation system [14]. In the presence of the bistability (figure 2c–e), the resulting cooperative state $X_2$ (for $k = n$, 100% cooperation C) is sustainable, even after the reward fund falls. Therefore, it is through the rise and fall of the reward fund that the coordination problem is completely resolved.

## 4. Discussion

Voluntary rewards can provide a powerful mechanism for overcoming coordination problems, without considering second-order punishment. This is an intriguing scenario that cannot easily be predicted from traditional models with voluntary punishment [16]. So far, for the evolution of cooperation with costly incentives, second-order freeloading has been a problematic ingredient, which should be defeated or suppressed [13,19,20]. The present model is in striking contrast to previous models and can complete 100% cooperation when second-order freeloading terminates the voluntary rewarders.

Collaborating results for transforming defectors into cooperators in coordination games have recently been obtained by considering optional participation [21,22] or through institutional punishment [12]. Optional participation can provide a simple but effective resolution for escaping the social trap [13,16]. In human societies, however, there are many issues to which people are inevitably required to be committed, such as nationality, religion, energy and environment. The present model focuses on such an unavoidable situation, and thus, players are forcibly admitted to games.

Institutional punishment crucially influences the establishment of a stable level of cooperation, but in large groups it may face a coordination problem in itself [7,23]. That is, it would be difficult for a single punisher to make such a large impact that activates a sanctioning system that covers the whole group. What about punishing those who make no contribution to institutional punishment? This triggers an infinite regression to the question: who pays for (higher-order) punishment? In contrast, a reward fund can rise in response to a single volunteer and then spread in a population of defectors.

All in all, it is not such a frustrating message that cooperation with reward funds is so powerful that it is more likely to start in the social trap than with institutional punishment. Voluntary rewarding is an efficient mechanism that allows for resolving coordination problems with minimal risk.

**Acknowledgements.** We thank Karl Sigmund and Voltaire Cang. This study was supported by grant RFP-12-21 from the Foundational Questions in Evolutionary Biology Fund.

**References**

1. Barrett, S. & Dannenberg, A. 2012 Climate negotiations under scientific uncertainty. *Proc. Natl Acad. Sci. USA* **109**, 17372–17376. (doi:10.1073/pnas.1208417109)
2. Kollock, P. 1998 Social dilemmas: the anatomy of cooperation. *Ann. Rev. Sociol.* **24**, 183–214.
3. Archetti, M. & Scheuring, I. 2012 Review: game theory of public goods in one-shot social dilemmas without assortment. *J. Theor. Biol.* **299**, 9–20. (doi:10.1016/j.jtbi.2011.06.018)

4. Centola, D. M. 2013 Homophily, networks, and critical mass: solving the start-up problem in large group collective action. *Ration. and Soc.* **25**, 3–40. (doi:10.1177/1043463112473734)

5. Boza, G. & Számadó, S. 2010 Beneficial laggards: multilevel selection, cooperative polymorphism and division of labour in threshold public good games. *BMC Evol. Biol.* **10**, 336. (doi:10.1186/1471-2148-10-336)

6. Archetti, M. & Scheuring, I. 2011 Coexistence of cooperation and defection in public goods games. *Evolution* **65**, 1140–1148. (doi:10.1111/j.1558-5646.2010.01185.x)

7. Boyd, R., Gintis, H. & Bowles, S. 2010 Coordinated punishment of defectors sustains cooperation and can proliferate when rare. *Science* **328**, 617–620. (doi:10.1126/science.1183665)

8. Hayashi, N., Ostrom, E., Walker, J. & Yamagishi, T. 1999 Reciprocity, trust, and the sense of control a cross-societal study. *Ration. and Soc.* **11**, 27–46. (doi:10.1177/104346399011001002)

9. Milinski, M., Sommerfeld, R. D., Krambeck, H. J., Reed, F. A. & Marotzke, J. 2008 The collective-risk social dilemma and the prevention of simulated dangerous climate change. *Proc. Natl Acad. Sci. USA* **105**, 2291–2294. (doi:10.1073/pnas.0709546105)

10. Santos, F. C. & Pacheco, J. M. 2011 Risk of collective failure provides an escape from the tragedy of the commons. *Proc. Natl Acad. Sci. USA* **108**, 10421–10425. (doi:10.1073/pnas.1015648108)

11. Archetti, M. 2009 Cooperation as a volunteer's dilemma and the strategy of conflict in public goods games. *J. Evol. Biol.* **22**, 2192–2200. (doi:10.1111/j.1420-9101.2009.01835.x)

12. Vasconcelos, V. V., Santos, F. C. & Pacheco, J. M. 2013 A bottom-up institutional approach to cooperative governance of risky commons. *Nature Clim. Change* 3, 797–801. (doi:10.1038/nclimate1927)

13. Sasaki, T., Brännström, Å., Dieckmann, U. & Sigmund, K. 2012 The take-it-or-leave-it option allows small penalties to overcome social dilemmas. *Proc. Natl Acad. Sci. USA* **109**, 1165–1169. (doi:10.1073/pnas.1115219109)

14. Hauert, C. 2010 Replicator dynamics of reward & reputation in public goods games. *J. Theor Boil.* **267**, 22–28. (doi:10.1016/j.jtbi.2010.08.009)

15. Chen, X., Gross, T. & Dieckmann, U. 2013 Shared rewarding overcomes defection traps in generalized volunteer's dilemmas. *J. Theor. Biol.* **335**, 13–21. (doi:10.1016/j.jtbi.2013.06.014)

16. Sigmund, K. 2010 *The Calculus of Selfishness*. Princeton: Princeton University Press.

17. Sasaki, T. & Unemi, T. 2011 Replicator dynamics in public goods games with reward funds. *J. Theor. Biol.* **287**, 109–114. (doi:10.1016/j.jtbi.2011.07.026)

18. Bach, L. A., Helvik, T. & Christiansen, F. B. 2006 The evolution of n-player cooperation—threshold games and ESS bifurcations. *J. Theor. Biol.* **238**, 426–434. (doi:10.1016/j.jtbi.2005.06.007)

19. Sigmund, K., De Silva, H., Traulsen, A. & Hauert, C. 2010 Social learning promotes institutions for governing the commons. *Nature* **466**, 861–863. (doi:10.1038/nature09203)

20. Perc, M. 2012 Sustainable institutionalized punishment requires elimination of second-order free-riders. *Sci. Rep.* **2**, 344. (doi:10.1038/srep00344)

21. Bchir, M. A. & Willinger, M. 2013 Does a membership fee foster successful public good provision? An experimental investigation of the provision of a step-level collective good. *Public Choice* **157**, 25–39. (doi:10.1007/s11127-012-9929-9)

22. Wu, T., Fu, F., Zhang, Y. & Wang, L. 2013 The increased risk of joint venture promotes social cooperation. *PLoS ONE* **8**, e63801. (doi:10.1371/journal.pone.0063801)

23. Raihani, N. J. & Bshary, R. 2011 The evolution of punishment in n-player games: a volunteer's dilemma. *Evolution* **65**, 2725–2728. (doi:10.1111/j.1558-5646.2011.01383.x)

**Figures and figure captions**

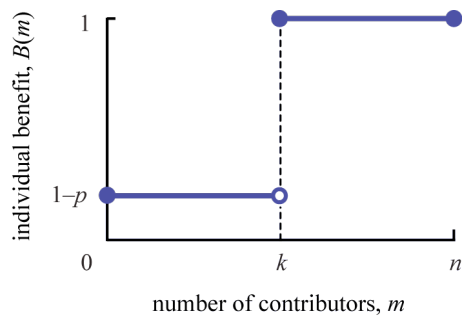

**Figure 1.** Step returns in the public good given by $B(m) = 1 - p$ for $m < k$, or otherwise $B(m) = 1$.

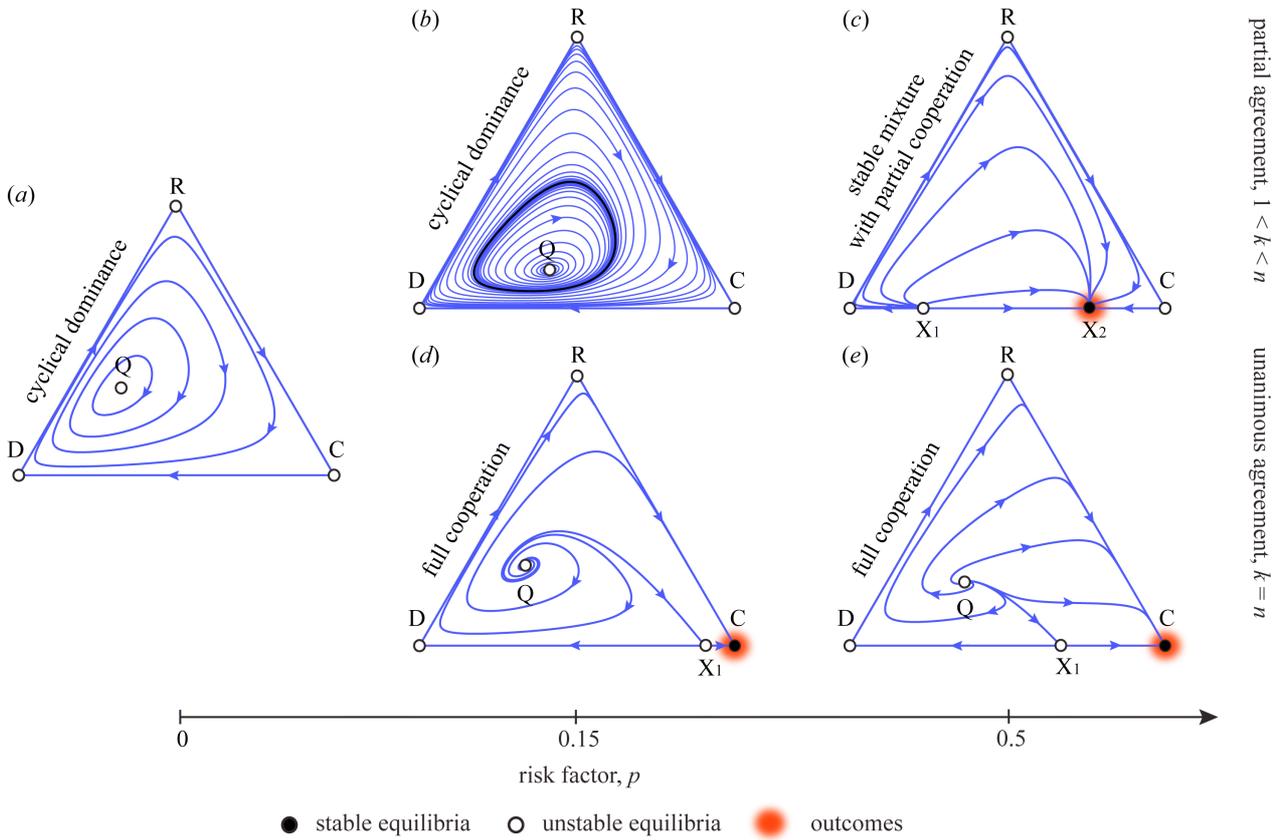

**Figure 2.** Threshold public good games with reward funds. Three corners of the state space, the node D: 100% defectors, the node C: 100% cooperators, and the node R: 100% rewarders are trivial equilibria. (*a*) Risk zero ($p = 0$). The unique interior equilibrium is a centre Q, which is neutrally stable, and is surrounded by closed orbits, which fill over the interior state space. Boundary orbits form a heteroclinic cycle. (*b*, *c*) Partial agreement ($1 < k < n$). In (*b*), for a small risk *p*, there can exist a stable limit cycle (bold black curve) along which three strategies dynamically coexist. In (*c*), when *p* goes beyond a critical value *p*\*, Q attains a point on the edge CD, from which then both unstable and stable equilibria sprout simultaneously; in particular, the stable one is a global attractor. (*d*, *e*) Unanimous agreement ($k = n$). When *p* increases beyond *p*\*, the unstable equilibrium enters the edge CD at the node C, which then turns into a sink, in particular, a global attractor. Parameters are $n = 5$, $c = 0.1$, $c' = 0.1$, $r' = 2.5$, and for (*b*) and (*c*), $k = 3$.

9